\documentstyle[11pt,aaspp4,epsf]{article}

\def\folio{\ifnum\pageno<2\nopagenumbers\else\number\pageno\fi}
\newtoks\headline \headline={\hss\twelverm\folio\hss} 
\newtoks\footline \footline={{\hfil}} 
\font\mathbf=cmmib10 scaled 1000             

\def\ref{\par\noindent\hangindent=2pc \hangafter=1 }
\def\amin{\ifmmode^{\prime}\else$^{\prime}$\fi}
\def\asec{\ifmmode^{\prime\prime}\else$^{\prime\prime}$\fi}

\def\cappage #1 #2 #3 {\vfill\eject\pageno=#1
\vglue 10 true in minus 10 true in \noindent{\bf Figure #2.} #3}
\def\ee #1 {\times 10^{#1}}   
\def\ut #1 #2 { \, \hbox{#1}^{#2}}
\def\u #1 { \, \hbox{#1}} 
\def\lsol{\, \hbox{$\hbox{L}_\odot$}}
\def\msol{\, \hbox{$\hbox{M}_\odot$}}
\def\kms {\, \hbox{km}\,\hbox{s}^{-1}}

\let\grad=\nabla
\def\cross{{\bf \times}}
\def\curl #1 {\grad \cross #1}
\def\div #1 {\grad \cdot #1}

\def\msol   {\hbox{$M_\odot$}}                  
\def\kms    {\hbox{km{\hskip0.1em}s$^{-1}$}}    

\begin{document}

\title{Proper Motion  of the Irradiated Jet HH 399 in the Trifid
Nebula}

\author{F. Yusef-Zadeh}
\affil{Department of Physics and Astronomy, Northwestern University,
Evanston, Il. 60208 (zadeh@northwestern.edu)}

\author{J. Biretta}
\affil{STScI, 3700 San Martin Drive, Baltimore, MD 21218
(biretta@stsci.edu)}

\author{M. Wardle}
\affil{Department of Physics, Macquarie University, NSW 2109, Australia
(wardle@physics.mq.edu.au)}

\begin{abstract}

HH 399 is one of the first Herbig Haro flows recognized to be irradiated
by the UV radiation of the massive O7.5 star in the Trifid nebula.  We
present the proper motion of the first irradiated jet based on two epochs
of HST observations of HH 399 separated nearly by five years using
H$\alpha$ and [SII] line filters.  High proper motion with continuous
velocities between 200$\pm$55 and 528$\pm24$ \kms are detected in both
lines along the 18$''$ extent of the jet axis.  The irradiated
fully-ionized jet consists  of  numerous  knots along the jet 
but also shows the
evidence for  a number of isolated blob-like structures running immediately
outside the jet with lower transverse velocities.  The transverse
velocities combined with radial velocity measurements indicate that the
jet axis lies away from the plane of the sky by only few degrees.  We
argue that the jet is fully  ionized based on [SII]/H$\alpha$ line ratio
as well as  radio continuum emission detected from the full extent of  the
jet at
3.6cm wavelength. The stellar mass-loss rate producing HH 399 is
estimated to be $\approx
2\times10^{-6} \msol $~yr$^{-1}$.

\end{abstract}

   
\section{Introduction}

The Trifid nebula (M20) is approximately 10$'$ in diameter
($\sim$8pc at a 
distance of 2.67 kpc)
 and is centered on a small cluster of hot stars
dominated by components A through G of HD 164492.  The ionizing flux
required to maintain the HII region, $10^{48.8}$ s$^{-1}$, is supplied
by the O7.5III star HD 164492A (e.g., Walborn 1973; Yusef-Zadeh et al.
2000), which has M$_v$=--5.3 for A$_v\approx$1.3 towards the central
stars (Lynds and O'Neil 1985). Several molecular condensations
associated with protostellar sources lie within the HII region
suggesting that a new generation of massive star formation has been
induced by the nebula (Cernicharo et al.\ 1998; Lefloch \& Cernicharo
2000).  The discovery of a remarkable optical jet associated with one of
the protostellar condensations embedded in the head of a cometary
globule was reported by Cernicharo et al.\ (1998). The {\it Hubble Space
Telescope} observations of this jet using H$\alpha$, SII, OIII lines
showed a number of knots along the 18$''$ extent of this continuous jet
(Hester et al. 1999).  This is the first jet that was reported in a
young H II region excited by the UV radiation from the central star, and
the first recognized to be propagating in a fully ionized medium (Hester
et al. 1997).  HH 399 is a member of a new class of irradiated  HH objects
that have been identified  in other star forming regions such as the Orion
and Pelican nebulae (Reipurth et al. 1998; Bally and Reipurth 2001, 2003). 
This implies that if the jet 
is  fully  ionized by the UV radiation field, the physical
characteristics of the outflow
such as the mass-loss rate can be measured directly. 

As part of constructing an HST mosaic image of the M20 nebula, we
measured proper motions of the knots in HH 399 between our
observations and those that had been made by Hester et al. (1999) nearly
five years earlier. In spite of different sensitivity and spatial
coverage between these two observations, we detect proper motion of as
much
as two WPC2 pixels along the jet during this interval. 
 The differing signal to noise ratio in the two  epoch data sets is not of 
concern because the noise is treated properly.
A shift of one
pixel ( 3.99$\times10^{15}$ cm at the distance of 2.67 kpc)  during this
time interval corresponds to 264.2 \kms.  The distance to M20 has been
measured by Lynds, Canzian \& O'Neil (1985) and Kohoutek et al (1999) to
be 1.67 and 2.67 kpc, respectively. Following Hester et al. (1999), we
adopt a distance of 2.67 kpc to HH 399. 
  Motivated by the low radial
velocity measurements along HH 399 (Rosado et al. 1999) and the
prediction that HH 399 should have a high transverse velocity
propagating into an ionized medium, we studied the proper motion of HH
399 using HST in H$\alpha$ and [SII] lines and combined them with
ground-based radial velocity measurements to determine the
three-dimensional velocity structure of this jet. This study shows a
high tranverse velocity of up to 500 $\kms$ along the jet axis lying
within few degrees of the plane of the sky.

\section{Observations and Data Reductions}

We completed WFPC2 observations (9104)  of the central region of M20 in
June 2002 (Day 173)  including its SE corner where HH 399 lies.  Our
spectral line observations with the F656N (H$\alpha$) and F673N ([SII])
filters overlapped earlier observations (6857) made by Hester et al.
(1999) in Sept. 1997 (Day 251).  A more detailed account of the large
scale study of ionized gas as well as the dust lanes of M20 will be given
elsewhere.  The exposure time per filter differed by a factor of 10 to 13
between in the first and second epoch observations. Thus, the 
uncertainties in
our present proper motion results are dominated entirely by the short 400
(400) sec exposure of the 2002 data compared to 4000 (5400) sec of the
1997 data for the H$\alpha$ ([SII]) lines. 
 Initially the images from each epoch were calibrated
and combined in the standard way to remove cosmic ray   
artifacts. We then used the STSDAS task WMOSAIC in IRAF  to remove
geometric distortion from the images.  
The images from the
two epochs
used  the IMLINTRAN task (i.e. rotated and shifted). 
For conveniece, the
parameters for this calculation were
chosen so as to place the jet along the X-axis of the pixel grid.
The first and second epoch images were 
rotated by --68$^0$ and -158$^0$ to align the long axis of the jet in the
E-W direction. 
The alignment of the images was checked using eight and nine stars in
H$\alpha$ and [SII] line data, respectively. 
The positions of the  stars which are distributed near
the jet are determined from  using IRAF IMCENTROID followed by small
adjustments
to
the IMLINTRAN input parameters for the 2001 data, and IMLINTRAN
was run again (always starting with the WMOSAIC output).
This process was iterated until the alignment error between the 
epochs was minimized.
For the final aligned images the eight stars had an
average position error of 0.01 pixel with a standard deviation
of about 0.13 pixel on each axis.  The uncertainty in the                      
alignment was thus judged to be 0.05 pixel.
This is much smaller than typical proper motions seen
in the jet (~1.5 pixels).  Importatly, any error in the
alignment would contribute an overall shift in the velocity
scale, rather than error in any one jet feature.

Hence this resulted in a pair of aligned images, each of which
had been re-sampled twice -- once to remove geometric distortion,
and once  for alignment. 
We  do not expect that there are any serious issues with re-sampling
the data.  Any such problems would most stongly impact unresolved
features.  For example, the 8 stars we used for alignment have 
a scatter of about 0.13 pixels RMS.  This probably sets an upper
limit for any re-sampling problems.  The jet features, which are
extended, should have much less re-sampling problems than the 
stars.   We  would expect re-sampling errors to scale inversely 
with the circumference around a feature.  So a jet blob with a scale of 
1" across will have about 1/10 the re-sampling error of a
star, or less than 0.01 or 0.02 pixel. We don't believe  re-sampling
is an important source of error.

To measure the two-dimensional transverse velocities, we
applied cross-correlation method of Biretta, Owen, \& Cornwell (1989).  
This method determines the fractional pixel position shift between images
(pixel size 99.6 milli-arcseconds) which maximizes their
cross-correlation.  Uncertainties on the velocities are estimated by
adding ten different noise images and measuring the dispersion of the
measured proper motions.  The noise images were selected from nearby
regions of the images containing only diffuse emission.

 Morphological changes in the jet are very minor, and
are less important than other effects (e.g. photon noise)
in the images.  We have generated an image where the
first epoch image is shifted forward according to the measured
velocities, and then subtracted from the second epoch image.
This image shows no evidence of the jet, and noise peaks
in the jet region are comparable in strength to those
visible throughout the image.

In order to determine if the HH 399 jet is fully ionized, we used the
archival data based on observations made with   
the Very Large Array (VLA) of the
National Radio Astronomy Observatory\footnote{The National Radio
Astronomy Observatory is a facility of the National Science Foundation,
operated under a cooperative agreement by Associated Universities, Inc.}. 
The observation was   carried out on March 13, 1998 in the A-array
configuration of the VLA at 8.4 GHz. Starndard calibration was done 
using the flux and phase calibrators before the final image 
was convolved to a resolution of 1$''$.

\section{Results}

Figure 1 shows the features in the vicinity of the jet based on a high
resolution WFPC2 H$\alpha$ image of the SE of the nebula. The emission
from the bright semi-circular rim is dominated by a photoionized
evaporative flow from the surface of the molecular cloud (Hester et al.\
1999; Cernicharo et al. 1998; Yusef-Zadeh et al. 2000; Lefloch et al.
2002). The jet as well
as an elephant trunk of cold gas protruding into the ionized medium are
also noted.  The jet consists of numerous equally spaced knots extending
over 18$''$ corresponding to a length of about 0.24 pc. Ground-based
observations indicate a bow shock structure 65$''$ (0.9 pc) away from
the base of the jet but aligned along the jet axis (Rosado et al. 1999).  
Detailed examination of H$\alpha$ emission in Figure 1 shows a weak and
diffuse emission continues beyond the edge of the bright jet. 
Extension of this weak emission terminates where the bow
shock is located. 
 This  diffuse H$\alpha$ emission
is at a  level above  the   background surface brightness 
of 5$\times10^{-16}$ erg cm$^{-3}$ s$^{-1}$ pixel$^{-1}$. 
This figure also
reveals a number of blobs located to the eastern side of the jet which
will be discussed below.

In order to measure the proper motion velocities, we selected boxes (A
-- M), as drawn on Figure 2 and listed in Table 1, in the direction
along and perpendicular to the jet axis.  The displacements in X and Y
directions (positive values to the west and north) during the time
interval between the two epochs and their corresponding transverse
velocities are shown for each box in columns 4 to 7.  
Column 8 shows the position angle ($\Theta$ = tan$^{-1} (V_y /V_x)$) of the
velocity vectors with respect to 
the jet axis using the [SII] proper motion data. The velocity vectors 
corresponding to each box have
also been 
graphically overlaid on  Figure 2. 
The radial
velocities of [SII] lines, as measured by Rosado et al. (1999), are
indicated in column 9. The flux of [SII] and H$\alpha$ line emission and
their intensity ratio [SII]/H$\alpha$ for each box are listed in columns
10 -- 12.  To measure the observed flux of H$\alpha$ and [SII] line
emission from the jet, the strong background contamination from the HII
nebula was measured locally off the jet axis before the background
emission
was
subtracted from the jet. The line ratio shown in the last columns
reflects the ratio of the background-subtracted [SII]/H$\alpha$ line
emission.

Comparison of data taken in mid-1997 (Hester et al. 1999) with our data
from mid-2002 show motions throughout the 18$''$ extent of the jet axis
both in H$\alpha$ and [SII] lines.  A qualitative presentation of the
proper motion is evidenced in Figure 4 by an image resulting from the
subtraction of the first and second epoch H$\alpha$ images from each
other. The proper motion is predominantly detected along the jet axis in
both [SII] and H$\alpha$ line images. Table 1 shows that the displacements
in H$\alpha$ and [SII] line emission along the jet axis are generally
consistent with each other within error bars. The velocities along the jet
range between 200 and 530 \kms with average velocities of $\approx$380
\kms (these values are reduced by 0.6 if the distance to HH 399 is
assumed to be 1.67 kpc). Figure 3 shows a plot of the velocities (V$_x$ and
V$_y$)
in both H$\alpha$ and [SII] lines represented in filled and open circles,
respectively. The vertical dotted line separates the boxes along the
length of the jet (A -- J) from those east of the jet (K -- M).  The
regions near the base of the jet tend to have lower velocities than in the
regions closer to the tip of the jet by $\approx$100 \kms. 
However, the boxes near the base of the jet, in particular Box J,  may be
confused by the 
emission from the rim which is likely to be stationary. 
The
highest proper motions or transverse velocities are detected in Box H3, as
drawn in Figure 2.  Box H3 shows the highest velocity of 528 \kms from a
region which is about 5$''$ from the base of the jet.  Marginal evidence
for deceleration with distance is seen from Box H to Box A (20\% reduction
in speed). The variation of
transverse velocities from the average velocity along the jet axis ranges
between 50 and 100 \kms, the latter of which is noted closer to the base
of the jet where there may be confusion due to stationary  features. 
There are no morphological
changes that we notice to first order. All
changes between the two epochs are
well-described as
motions.  Any other changes are lost in the noise.  
The features are not simple bullets, but the timescale for
significant structural change appears to be longer than our time  
baseline, at least for our signal-to-noise ratio.

There is also 
3$\sigma$ detection of 
transverse motions perpendicular to the jet axis. We note that all
the measured  proper motion or transverse velocities normal to 
the jet axis having 1--3$\sigma$ detections 
have positive values  (oriented to the north). 
In particular, Boxes D, F, H and I
show vertical velocities  ranging between 26 and 77 \kms in the
direction normal to  the jet axis, as noted in the bottom
panel
of Figure 3.  
A more significant tranverse motion away from 
the jet axis is noted along three  isolated blobs  (K -- M) of ionized
gas.  The  proper motion velocities of these blobs
indicate that they are part of the outflow but differences are noted 
in proper motion velocities in   H$\alpha$ and [SII] lines.
In contrast to the ratio
of the tranverse velocities along the jet  
which is V$_x$/V$_y\sim10$, this 
ratio is $\approx2-3$  for the blobs.

Figure 5a shows contours of background-subtracted H$\alpha$ line
emission whereas Figure 5b shows the distribution of ratio of
[SII]/H$\alpha$ line emission.  
We note several faint features extending
from the face of the jet directed to the northeast. These
"tails" 
of emission are aligned along an axis that points to the bright
cluster of stars at the center of the
nebula. The surface brightness of H$\alpha$
emission from these faint 
``tails'' of ionized gas is estimated to be between 0.4 and
1$\times10^{-16}$ ergs
s$^{-1}$ cm$^{-2}$ pixel$^{-1}$.
The distribution of surface brightness ratio of [SII] to
H$\alpha$ line emission, as listed in Table 1, indicates a constant
value ranging between 0.19 and 0.36.  This ratio ranges between 0.13 and
0.15 for the background emission and 0.24 and 0.28 for the bright
ionized rim at the surface of the molecular condensation.  
Radio
continuum observations of the bright rim TC2  have shown the evidence of
free-free emission implying that the jet, having similar
[SII]/H$\alpha$ line ratio to that of the bright rim, is also fully
ionized along its axis (Yusef-Zadeh et al. 2000). 
In addition, the distribution of free-free
emission at
3.6cm, as shown in Figure 6, inidcates the presence of a continuous 
structure along the full extent of HH 399 and 
a radio morphology similar to that of its  optical counterpart.  
The grayscale and contour images of  the jet and the TC2 rim are 
shown 
with a resolution of 1$''$. The typical surface brightness of the 
jet at 3.6cm is  about 50 -- 75 $\mu$Jy beam$^{-1}$ which  corresponds to an 
emission measure of 2 -- 3 $\times10^4$  cm$^{-6}$ pc assuming that the
electron temperature is between 5$\times10^3$ to 10$^4$K. The electron
density n$_e = 1.5-3
\times 10^3 $
cm$^{-3}$
is estimated assuming that the diameter of the jet is between
2-4 $\times 10^{16}$
cm. 
These values are  consistent with other estimates of the density of ionized
gas in the jet (Lefloch et al. 2002; Rosado et al. 1999). 


\section{Discussion}

The proper motion measurements along the jet give a clear indication for
relatively uniform supersonic motion with velocities ranging between 300
and 500 \kms.  Ground-based radial velocity measurements in [SII] line
of HH 399 show peak radial velocities of $\approx$15 \kms increasing by
few \kms\ from the base of the jet to the tip of the jet.  Comparing the
radial and tranverse velocity measurements along the jet for Boxes A to
F suggests that the jet axis is tilted by $\approx$2-4$^0$ away from the
plane of the sky which is consistent with earlier predictions
(Cernicharo et al. 1998;  Rosado et al. 1999;  Hester et al. 1999).

The emission from the jet arises mostly from lines probing shocked
material propagating from the head of  a molecular condensation in the 
direction toward  the
fully ionized medium of the HII region. High velocity broad wings of
HCO$^+$ emission from molecular condensation (TC2 in Cernicharo et al.
1998) indicates that the jet is arising from a protostar with an age of
$\approx 10^4$ years.  Using a flow velocity of 400 \kms 
along the full 18$''$ extent of the HH 399 jet, 
a dynamical  time scale of $\approx$500 years is found. 
This age is about four times lower than the flow time scale to the bow
shock assuming the same jet velocity. Thus, the estimated time scales of 
HH flows appear to be much less than  the estimated age of the
molecular condensation TC2 from which the jet 
is launched. 

We believe the assumption of fully ionized jet by the external radiation
field is consistent with the observations, as discussed earlier.  The
mass-loss rate of the star can be estimated by 1.4 $\times\pi v_j r^2_j
n_j m_H$ where $v_j=400 \kms$, $n_j=10^3 cm^{-3}$ and
$r_j=6.5\times10^{-3} pc$ are the velocity, the ionized density and the
radius of the jet corresponding to a 0.5$''$, respectively.  We believe
the assumption of fully ionized gas applies better to the region closer
to the tip of the jet where [SII]-to-H$\alpha$ surface brightness ratio
is slightly lower than in the region near the base of the jet (Table 1).  
The electron density is estimated to be $\approx10^3$ cm$^{-3}$ from
H$\alpha$ intensity of the knots.
 The mass-loss rate is then estimated
to be $\approx 2\times10^{-6} \msol $ yr$^{-1}$. Given the
irradiated and fully ionized nature of the jet, we believe this is one
of the most 
direct estimates of the mass-loss rate of a young stelar 
object.

There are three reasons to consider the interaction hypothesis between the
jet, the ionized medium of the HII region and the radiation field of the
central cluster. One is the positive vertical motion noted on the face of
the jet normal to the jet axis.  Although most of the gas in the jet show
upper limit velocity of about 40 \kms in the direction normal to the jet
axis, Boxes H and I show vertical velocities of up to 77 \kms.  The other
is the evidence for three isolated blobs of ionized gas running on the
eastern side of the jet. Lastly, the faint tails of H$\alpha$ and [SII]
line emission, as see in Figure 5, are pointed toward east where the
central cluster of stars lies.

Rosado et al. (1999) found two bright and faint velocity components
associated with the jet. Neither of these components are spatially
resolved.  The brightest component with an electron density of n$_e\approx
10^2$ cm$^{-3}$ shows velocities between +25 and +30 $\kms$ which are
slightly redshifted, relative to the radial velocity of the background
nebulosity 19 $\kms$.  On the other hand, the fainter, broader component
with n$_e\approx 10^3$ cm$^{-3}$ is blueshifted by few \kms with respect
to the ambient medium.  The former density is similar to the density of
the ionized medium in the nebula.  The density of ionized gas of the
knots is consistent with the density of the bright
kinematic component. This implies that the low-density ionized gas of the
nebula is acclerated at the edge of the jet as the jet propagates in the
direction away from the HII region.  It is plausible to consider whether
the acceleration by radiation field of the cluster of stars or the ram
pressure of the ionized gas in the nebula is significant over the age of
the jet.  Acceleration by radiation pressure from O7.5III star is
estimated to change the velocity of the jet by 0.2 \kms whereas the
acceleration by the ram pressure in a wind is estimated to be 2 \kms over
the age of the jet. These estimates assumed that the distance between HD
164492 A with a luminosity of 10$^{5.41}$ \lsol (Panagia 1973) and the jet
is 1.7 pc and the wind number density and velocity are 100 cm$^{-2}$ and
20 \kms, respectively. It is clear that the radiation field is
insignificant in disturbing the material associated with the jet in spite
of the assumption that all of the radiation incident on the jet is
absorbed. If the ambient gas is associated with the jet material, the ram
pressure of the wind can accelerate the low density component of ionzied
gas associated with the jet by few \kms. These estimates, however, can not
explain the origin of blobs detected off the jet axis.

\section{Conclusions}

In summary, the irradiated jet HH 399 in the Trifid nebula shows a high
proper motion along its jet axis. Its transverse proper motion velocity
which is about 400 \kms dominates over its radial or vertical motion by
about an order of magnitude. The positive vertical  motion indicates 
that the jet is asymmetric in its motion with respect to
 the jet axis. Other asymmetries are noted in the distribtion of several
blobs and  long  "tails"
of H$\alpha$ and [SII] line emission from  the side of the jet where vertical
motion is detected. 
These morphological and kinematic asymmetries as well as a continuous
structure of  HH 399 are
distinguishing characteristics  of a remarkable  irrediated jet when
compared other jets such as HH 34 and HH 110 (Reipurth et al. 2002; Hartigan
et  al. 2001). 

The proper motion results suggest that to first
order  
HH 399 is moving as a unit as a number of  other  HH objects show similar 
trend in their motion
(e.g., Hatiagn 2003). 
There appears to be faint emission beyond its
18$''$ extent along the jet axis toward the direction of the bow shock.
This implies that the jet is continuous and the outflow from the central
star increased its density about 500 years ago.  We note that [SII]
emission from the jet with respect to H$\alpha$ emission does not shine
brilliantly suggesting that much of the gas along the jet is fully ionized
by the radiation field. This implies that much of the material in the jet
is detected and therefore we determine the velocity distribution of the
gas and the mass-loss rate accurately. Since the gas in the jet is fully
ionized, the clumpiness of the jet is also inferred to be due to density
enhancement.
With the exception of its asymmetric distribution, the "tails" of ionized gas
on one side of the HH 399 jet resembles the 
wings  of bow shocks noted in HH objects such as the HH 111 jet (Reipurth et
al. 1997). Furthermore, 
the origin of the blobs outside the jet and the faint "tails" are  not well
understood but we speculate  that they may be part of a low-density broad
outflow from the central star with the bright jet at its center or due to an 
axisymmetric distribution of the radiation field of the central
cluster of hot stars causing an instability 
along the jet axis (Ryutov et al. 2003).

\section{References}

\noindent
Bally, J. \& Reipurth 2002, AJ, 126, 893

\noindent
Hartigan, P.M. 2003, RevMexAA, 15, 112

\noindent
Hartigan, P., Morse, J.A., Reipurth, B., Heathcore, S. \& Bally, J. 
2001, ApJ, 559, L157

\noindent
Reipurth, B., Hartigan, P., Heathcore, S., Morse, J.  \& Bally,
J. 1997, 
ApJ, 114, 757

\noindent
Reipurth, B., Heathcore, S., Morse, J., Hartigan, P.  \& Bally, J. 2002,
ApJ, 123, 362

\noindent
Reipurth, B. \& Bally, J. 2001, ARA\&A, 39, 403

\noindent
Reipurth, B., Bally, J., Fesen, R. \& Devine, D. 1998, 
Nature, 396, 343 

\noindent
Biretta, J., Owen, F., \& Cornwell, T. 1989, ApJ,  
342, 128

\noindent
Cernicharo, J., Lefloch, B., Cox, P.,  Cesarsky, D.,  Esteban, C.,
Yusef-Zadeh, F.,  Mendez, D.I.,  Acosta-Pulido, J.,  Lopez, R. J.
Garcia,  \& Heras, A. 1998, Science, 282, 462

\noindent
Hester, J., Scowen, P., Stappelfeldt, K. \& Krist, J.
1999, BAAS,  194, 68.10

\noindent
Kohoutek, L.,  Mayer, P. \& Lorenz, R.
1999, A\&AS, 134, 129

\noindent
Lefloch, B. \& Cernicharo, J. 2000, ApJ, 545, 340

\noindent
Lefloch, B.,  Cernicharo, J., Rodriguez, L.F., Miville-Deschenes, M.A., 
Cesarski, D. \& Heras, A.  
2002, ApJ, 581, 335

\noindent
Lynds, B.T. and O'Neil, E.J. Jr. 1985, ApJ., 294, 578

\noindent
Lynds, B.T., Canzian, B.J. \& O' Neil, E.J., Jr.
1985, ApJ, 288, 164

\noindent
Panagia, N. 1973, AJ, 78, 929




\noindent
Rosado, M., Esteban, C., Lefloch, B., Cernicharo, J. \& Garcia Lopez, R.J.
1999, ApJ, 118, 2962

\noindent
Ryutov, D.D., Kane, J.O., Pound, M.W. \& Remington, B.A. 2003, Plasma Physics
Control Fusion, 45, 769

\noindent
Walborn, N.R. 1973, AJ, 78, 1067

\noindent
Yusef-Zadeh, F., Shure, M., Wardle, M. \& Kassim, N.   2000, ApJ,
540, 849 


%
%
\def\x      {$\times$}                          
\def\deg    {$^{\circ}$}                        
\def\amin   {$^{\prime}$}                       
\def\hour   {$^{\rm h}$}                        
\def\min    {$^{\rm m}$}                        
\def\dasec  {\hbox{$.\!\!^{\prime\prime}$}}     
\def\dsec   {\hbox{$.\!\!^{\rm s}$}}            
\def\hh     {\hbox{H$_2$}}                      
\def\thco   {\hbox{$^{13}$CO}}                  
\def\percmcub {\hbox{cm$^{-3}$}}                
\def\percmsq {\hbox{cm$^{-2}$}}                 
\def\kms    {\hbox{km{\hskip0.1em}s$^{-1}$}}    
\def\msol   {\hbox{$M_\odot$}}                  
\def\tex    {\hbox{$T_{ex}$}}                   


{\tiny
\begin{deluxetable}{lcccccccccc}
\tablenum{1}
\tablecolumns{11}
\tablewidth{0pt}
\tablecaption{Velocities of gas along the HH 399 jet}
\tablehead{
\colhead{Box} & 
\colhead{Size} &
\colhead{Filter} &
\colhead{X$_{shift}$} &
\colhead{Y$_{shift}$} &
\colhead{$V_{\rm x}$} &
\colhead{$V_{\rm y}$} &
\colhead{$\Theta$} &
\colhead{$V_{\rm z}^a$} &
\colhead{Flux (10$^{-14}$)} &
\colhead{[SII]/} 
\cr
\colhead{Num.} &
\colhead{($\Delta{\rm x}\times\Delta{\rm y}$)} & 
\colhead{} &
\colhead{(pixel$^b$)} &
\colhead{(pixel$^b$)} &
\colhead{(\kms)} &
\colhead{(\kms)} &
\colhead{($\circ$)} &
\colhead{(\kms)} &
\colhead{(ergs s$^{-1}$)} &
\colhead{H$\alpha$} 
\cr
\colhead{} &
\colhead{(asec$\times$asec)} & 
\colhead{} &
\colhead{} &
\colhead{} &
\colhead{} &
\colhead{} &
\colhead{} &
\colhead{} &
\colhead{(cm$^{-2}$)} &
\colhead{} 
\cr
}
\startdata

A & 1.69$\times$1.69 &   [SII]& 1.38$\pm0.07$& 0.13$\pm$0.09&364$\pm$18 
&34$\pm$23 & 5.5& 17   & 0.96 & 0.19  \cr
&                    & H$\alpha$ &  1.43$\pm$0.05 & 0.07$\pm0.03$   &
378$\pm$13 & 18$\pm$8 &2.7  & 5.12 &   \cr

B & 1.69$\times$1.20 &   [SII]& 1.39$\pm0.47$ & 0.21$\pm$0.21&    
367 $\pm$124 & 55$\pm$55   &8.5& 15 & 0.56 & 0.26   \cr
&                    & H$\alpha$ &  1.76$\pm$0.20 & 0.12$\pm0.06$   &
465$\pm$52 & 32$\pm16$   &3.9&  & 2.10 &   \cr

C & 1.59$\times$1.59 &   [SII]& 1.45$\pm0.14$ & 0.18$\pm$0.07&    
383$\pm$37 & 47$\pm$18   &7&  14 & 1.10 & 0.24   \cr
&                    & H$\alpha$ &  1.55$\pm$0.14 & 0.08$\pm0.10$   &
 409$\pm$37 & 13$\pm$26   &1.8&  & 4.55 & \cr

D & 1.59$\times$1.59 &   [SII]& 1.50$\pm0.18$ & 0.15$\pm$0.05&    
 396$\pm$47 & 40$\pm$13   &5.8&  15  & 1.66 & 0.30  \cr
&                    & H$\alpha$ &  1.36$\pm$0.18 & 0.1$\pm0.03$   &
 360$\pm$48 & 26$\pm$8   &4.1&   & 5.29 &   \cr

E & 1.59$\times$1.59 &   [SII]& 1.37$\pm0.20$ & 0.16$\pm$0.08&    
362 $\pm$53 & 42$\pm$21   &6.6& 13  & 1.89 & 0.28  \cr
&                    & H$\alpha$ &  1.38$\pm$0.12 & 0.08$\pm0.04$   &
 365$\pm$32 &21$\pm$11   &3.3&   & 6.59 &   \cr

F & 1.59$\times$1.59 &   [SII]& 1.65$\pm0.16$ & 0.21$\pm$0.18&    
 436$\pm$42 & 55$\pm$48   &7.2& 13  & 1.73 & 0.28  \cr
&                    & H$\alpha$ &  1.47$\pm$0.12 & 0.20$\pm0.04$   &
 388$\pm$32 & 53$\pm$11   &7.8&   & 6.14 &  \cr

\cr

H & 1.20$\times$1.20 &   [SII]& 1.66$\pm0.20$ & 0.29$\pm$0.05&    
 439$\pm$53 & 77$\pm$13   &9.9&  & 1.34 & 0.36  \cr
&                    & H$\alpha$ &  1.92$\pm$0.03 & 0.21$\pm0.05$   &
  507$\pm$8 & 55$\pm$13   &6.2&   & 3.68 &  \cr

I & 1.10$\times$1.10 &   [SII]& 1.28$\pm0.24$ & 0.29$\pm$0.10&    
 338$\pm$63 & 77$\pm$26   &12.8&  & 1.0 & 0.30   \cr
&                    & H$\alpha$ &  1.04$\pm$0.17 & 0.14$\pm$0.02   &
  275$\pm$45 & 37$\pm$11   &7.7&   & 3.3 &  \cr

J & 1.10$\times$1.10 &   [SII]& 0.84$\pm0.36$ & 0.13$\pm$0.22&    
 221$\pm$95 & 34$\pm$58   &8.7&   & 0.66 & 0.30   \cr
&                    & H$\alpha$ &  0.77$\pm$0.21 & 0.13$\pm0.11$   &
  203$\pm$55 & 34$\pm$29   &9.5&  & 2.19 &   \cr
\cr
H1 & 0.7$\times$0.5   & [SII] &  1.37$\pm$0.40 &
0.11$\pm0.12$ & 362$\pm$106 & 29$\pm$32   &4.6  & 0.58 & 0.33   \cr
 &   & H$\alpha$ &  1.04$\pm$0.07 &
0.02$\pm0.06$ & 275$\pm$18 & 5$\pm$16&1   &  & 1.76  \cr

H2& 0.5$\times$0.8  & [SII] &  --- &
--- & ---
   & ---  &---&   &0.58 &0.36   \cr
& & H$\alpha$ &  1.76$\pm$0.07 &
0.22$\pm0.06$ &
  465$\pm$18 & 58$\pm$16   &7.1  & 1.59 &  \cr

H3& 0.7$\times$0.6  & [SII] &  1.77$\pm$0.30 &
0.59$\pm0.37$ &
468 $\pm$79 & 156$\pm$98   &18.4  &0.60 & 0.34   \cr
& & H$\alpha$ &  2.0$\pm$0.09 &
0.16$\pm0.09$ &
528 $\pm$24 & 42$\pm$24   &4.5&  & 1.76 &  \cr

\cr

K & 1.20$\times$1.29 &   [SII]& 0.08$\pm0.72$ & 0.74$\pm$0.23&    
 21$\pm$190 & 195$\pm$61   &84&  &0.13 & 0.21   \cr
&                    & H$\alpha$ &  1.04$\pm$0.21 & 0.08$\pm0.25$   &
  275$\pm$55 & 21$\pm$66   &4.3  & 0.62 &  \cr

L & 0.9$\times$0.9 &   [SII]& 0.80$\pm0.28$ & 0.37$\pm$0.31&    
 211$\pm$74 & 98$\pm$82   &25&  & 0.15 & 0.23  \cr
&                    & H$\alpha$ &  1.17$\pm$0.25 & 0.72$\pm$0.20   &
 309$\pm$66 & 190$\pm$53   &31.6&  & 0.65 &   \cr

M & 0.8$\times$0.8 &   [SII]& 0.75$\pm0.27$ & 0.51$\pm$0.27&    
 198$\pm$71 & 135$\pm$71   &34&  & 0.12 & 0.25   \cr
&                    & H$\alpha$ &  0.54$\pm$0.21 & 0.99$\pm0.28$   &
  143$\pm$56 & 262$\pm$74   &61&  & 0.48 &   \cr
\cr

\enddata
\tablenotetext{\it a}{Radial velocity from Rosado et al. (1999)}
\tablenotetext{\it b}{0.09965 asec}
\end{deluxetable}
}


\vspace{1.0 truecm}
\begin{figure}
\plotfiddle{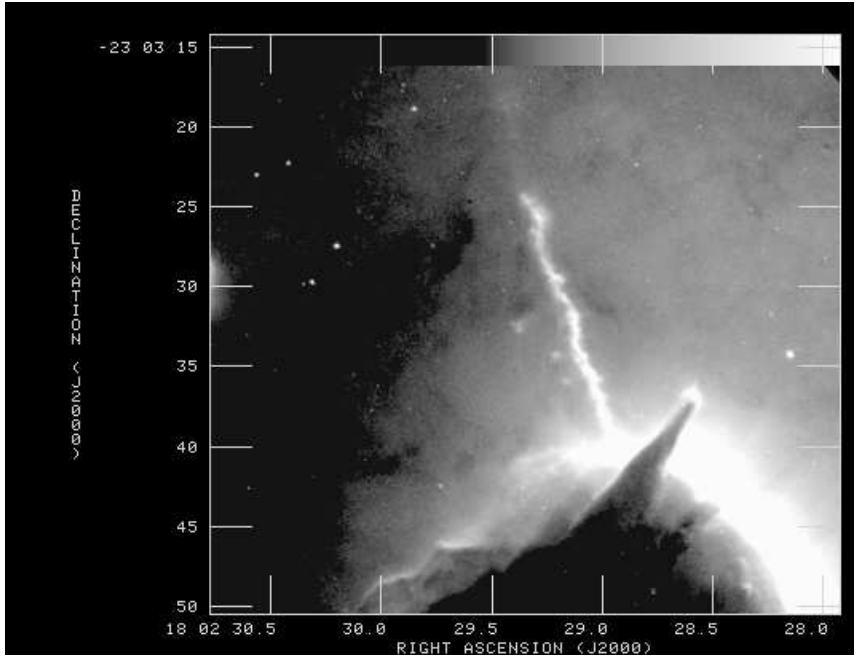}{2.5in}{0}{60.0}{60.0}{-200}{-125}
\caption{A large scale distribution of  H$\alpha$ emission from
the 1997 epoch showing the HH 399 jet propagating into a low density 
ionized medium associated with the Trifid nebula. The central stellar
cluster  responsible for the ionization of the jet lies to the northwest.}
\end{figure}

\vspace{1.0 truecm}
\begin{figure}
\plotone{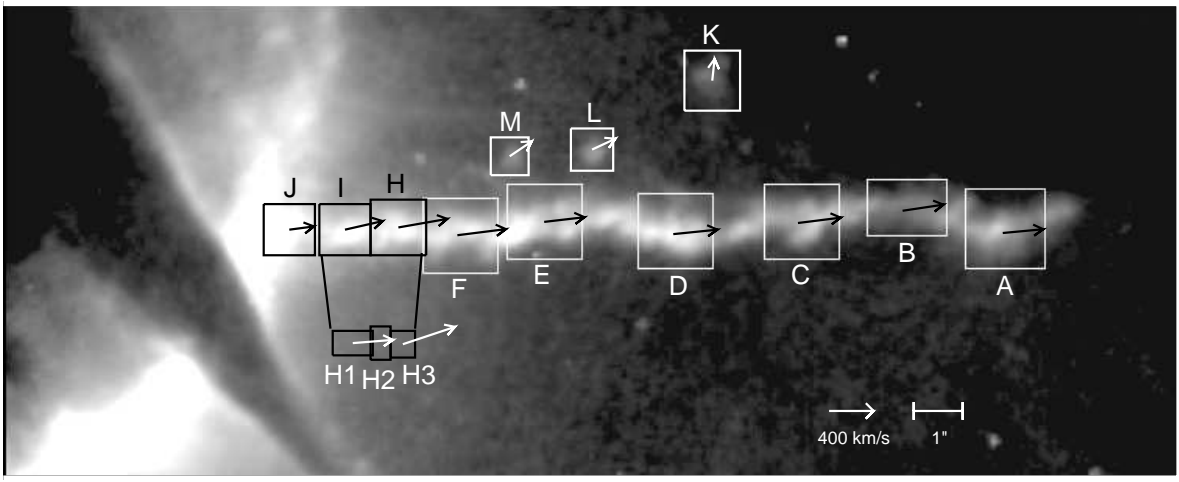}
\caption{An [SII]  image of HH 399 from the 1997 epoch and the drawn
boxes for 
extracting proper
motion  measurements listed in Table 1. Boxes HI, H2 and H3 show
the regions where the highest tranverse velocity in Box H and I  has  been
noted. The lines drawn on each box shows the transverse velocity vector
based on [SII] proper  motion data. The position angle ($\Theta$) of  each
velocity
vector is measured from tan$^{-1} (V_x / V_y)$, as listed in Table 1 }. 
\end{figure}

\vspace{1.0 truecm}
\begin{figure}
\plotfiddle{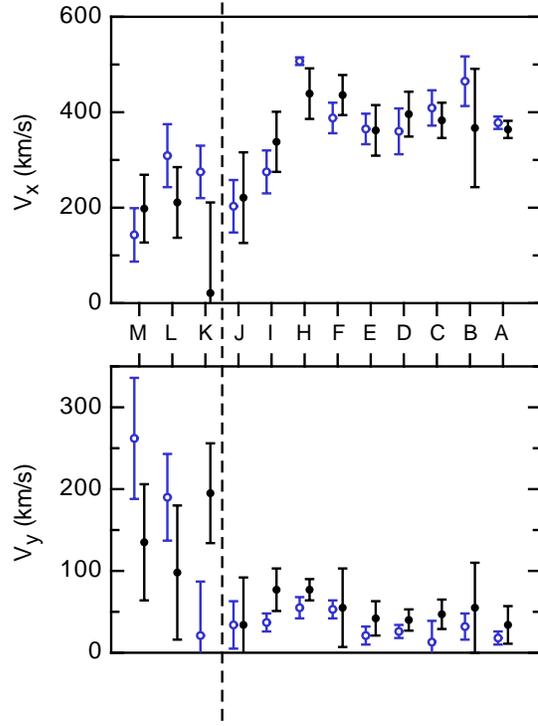}{2.5in}{0}{90.0}{90.0}{-200}{-100}
\caption{ The top and bottom  panels show  proper motion velocities
along and normal  to the jet axis, respectively. The filled and empty 
circles with their 1-$\sigma$ error bars represent H$\alpha$ and [SII]
lines, respectively. The dotted
line
separates the velocities of  the boxes along the jet axis from those
boxes, K through M,  as  noted in Figure 2. }
\end{figure}

\vspace{1.0 truecm}
\begin{figure}
\plotfiddle{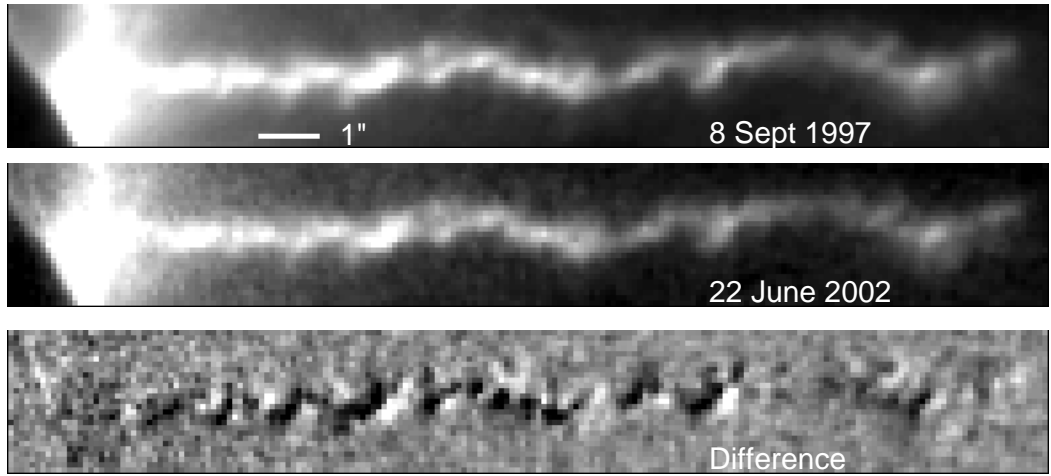}{2.5in}{-90}{70.0}{70.0}{-270}{340}
\caption{ The top and middle panels show  the first and second epoch 
WFPC2 H$\alpha$ images of
the HH 399
jet, respectively.  The bottom panel displays  the difference image made
by subtracting the top image from the middle image.}
\end{figure}

\vspace{1.0 truecm}
\begin{figure}
\plotfiddle{f5a.ps}{2.5in}{-90}{50.0}{50.0}{-200}{350}
\plotfiddle{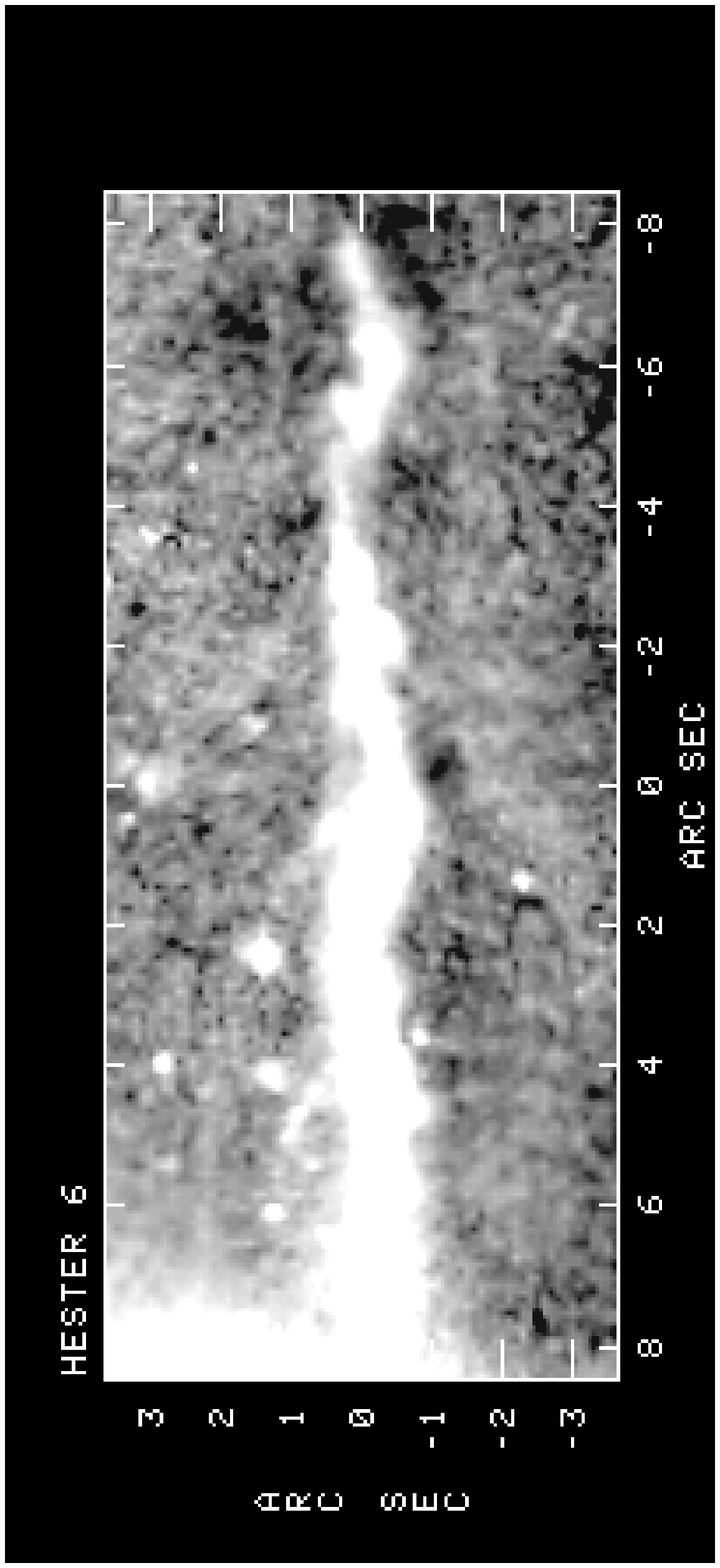}{2.5in}{-90}{50.0}{50.0}{-200}{300}
\caption{The top panel shows the distribution of background subtracted  
H$\alpha$ emission. The bottom panel shows the
ratio 
of the surface brightness of [SII] to H$\alpha$ line emission.}
\end{figure}

\vspace{1.0 truecm}
\begin{figure}
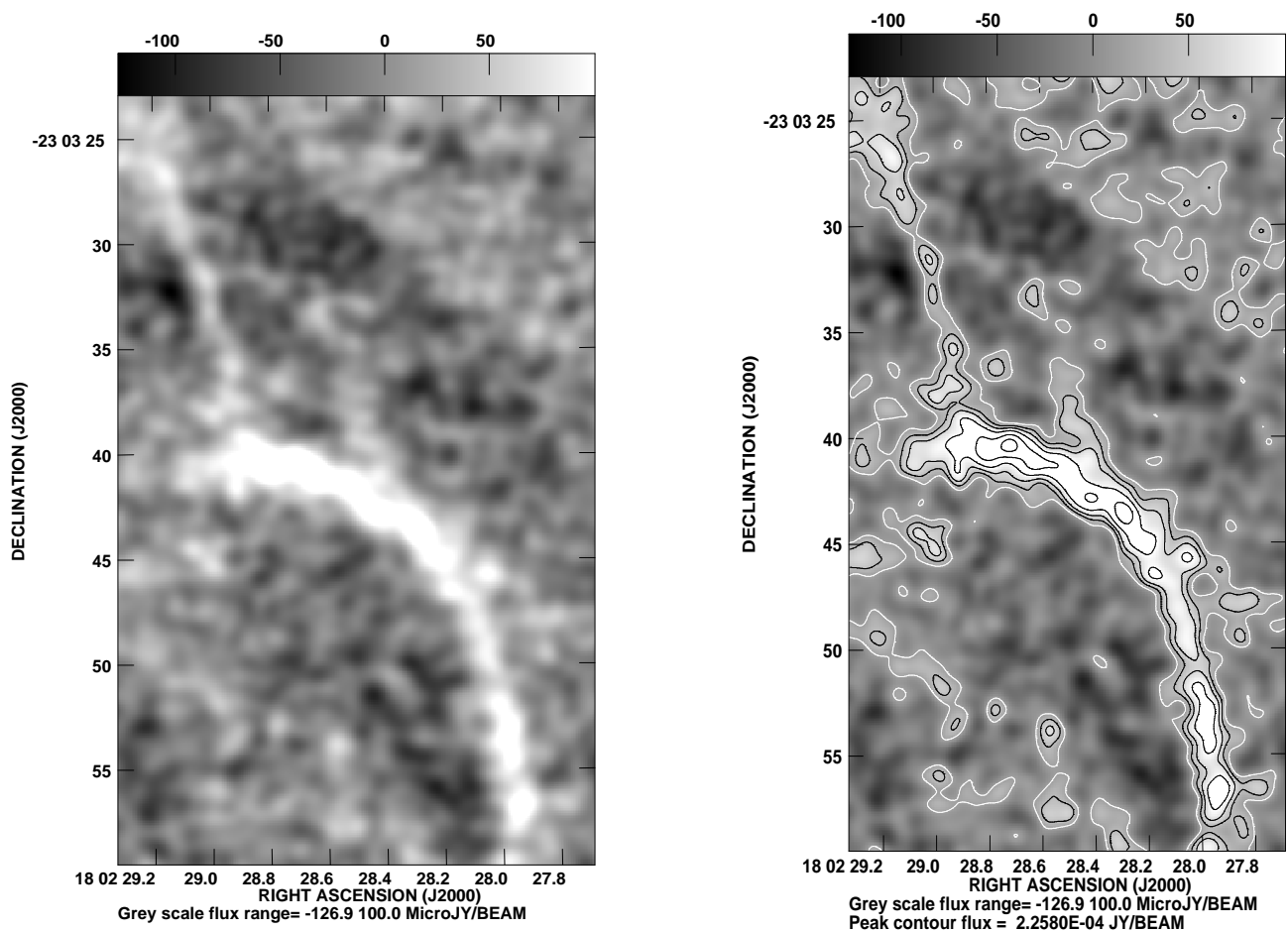

\plotfiddle{f6a.ps}{2.5in}{0}{50.0}{50.0}{-300}{-150}
\plotfiddle{f6b.ps}{2.5in}{0}{50.0}{55.0}{-30}{25}
\caption{The left panel shows the grayscale distribution of 3.6cm
continuum 
emission  from the ionized TC2 rim and HH 399 whereas the right panel
shows the contour representation of the same region  with levels 
0.3, 0.8, 1.3, 2, 3, 4, 5 $\times$ 50 $\mu$Jy beam$^{-1}$ with a
resolution of 1$''^2$. The rm noise is 20 $\mu$Jy beam$^{-1}$.}
\end{figure}
\end{document}